\documentclass[aps,prc,twocolumn,superscriptaddress,nofootinbib]{revtex4}
\usepackage{graphicx,amsfonts,color,epsfig,epstopdf}
\usepackage{footnote,multirow}
\usepackage{amsmath}
\usepackage{amssymb,bbold}
\usepackage{physics}
\usepackage[utf8]{inputenc}

\newcommand{\SU}[1]{\ensuremath{\mathrm{SU}( #1 )}}

\newcommand{\SpR}[1]{\ensuremath{\mathrm{Sp}( #1,\mathbb{R} )}}
\newcommand{\ad}{\ensuremath{ a^\dagger }}

\newcommand{\RedME}[3]{\ensuremath{\langle #1 \| #2 \| #3 \rangle}}

\newcommand{\eg}{\emph{e.g.}}
\newcommand{\etal}{\emph{et al.}}

\newcommand{\Nmax}{$N_\mathrm{max}$}

\newcommand{\NNLOopt}{NNLO$_\mathrm{opt}$}
\newcommand{\Li}[1]{\ensuremath{ ^{#1}\mathrm{Li}}}
\newcommand{\hw}{$\hbar\Omega$}

\begin{document}
\title{\textit{Ab initio} single-neutron spectroscopic overlaps in lithium isotopes}
\author{G. H. Sargsyan}
\affiliation{Department of Physics and Astronomy, Louisiana State University, Baton Rouge, LA 70803, USA}
\affiliation{Lawrence Livermore National  Laboratory, Livermore, California 94550, USA}
\author{K. D. Launey}
\affiliation{Department of Physics and Astronomy, Louisiana State University, Baton Rouge, LA 70803, USA}
\author{ R. M. Shaffer}
\affiliation{Department of Physics and Astronomy, Louisiana State University, Baton Rouge, LA 70803, USA}
\author{ S. T. Marley}
\affiliation{Department of Physics and Astronomy, Louisiana State University, Baton Rouge, LA 70803, USA}
\author{N. Dudeck}
\affiliation{Department of Physics and Astronomy, Louisiana State University, Baton Rouge, LA 70803, USA}
\affiliation{School of Physics, Georgia Institute of Technology, Atlanta, Georgia 30332, USA}
\author{A. Mercenne}
\affiliation{Center for Theoretical Physics, Sloane Physics Laboratory, Yale University, New Haven, Connecticut 06520, USA}
\affiliation{Department of Physics and Astronomy, Louisiana State University, Baton Rouge, LA 70803, USA}
\author{T. Dytrych}
\affiliation{Department of Physics and Astronomy, Louisiana State University, Baton Rouge, LA 70803, USA}
\affiliation{Nuclear Physics Institute of the Czech Academy of Sciences, 250 68 \v{R}e\v{z}, Czech Republic}
\author{J. P. Draayer}
\affiliation{Department of Physics and Astronomy, Louisiana State University, Baton Rouge, LA 70803, USA}

\begin{abstract}
We calculate single-neutron spectroscopic overlaps for lithium isotopes in the framework of the \textit{ab initio} symmetry-adapted no-core shell model. 
We report the associated neutron-nucleus asymptotic normalization coefficients (ANCs) and spectroscopic factors (SFs) that are important ingredients in many reaction cross section calculations. While spectroscopic factors have been traditionally extracted from experimental cross sections, their sensitivity on the type of reactions, energy, and the underlying models point to the need for determining SF from first-principle structure considerations. As illustrative examples, we present $^6$Li+n, $^7$Li+n, and $^8$Li+n, and we show that the results are in a good agreement with those of other \textit{ab initio} methods, where available, including the quantum Monte Carlo approach. We compare ANCs and SFs to available experimentally deduced values, with a view toward expanding this study to heavier nuclei and to extracting inter-cluster effective interactions for input into analyses of existing and future experimental data.


\end{abstract}

\maketitle
\section{Introduction}

In the recent years there has been a significant interest in describing nuclear reactions from \emph{ab initio} approaches, 
and especially in constructing from first principles effective inter-cluster interactions, often referred to as optical potentials. 
Various methods have been developed to derive \emph{ab initio} optical potentials including 
 the Green’s function method with Coupled-Cluster many-body calculations \cite{RotureauDHNP17, Rotureau_2020}, the self-consistent Green’s function framework \cite{idini19}, the multiple scattering approach for intermediate projectile energies \cite{BurrowsEWLMNP19,BurrowsBEWLMP20,VorabbiGFGNM2022}, from two- and three-nucleon chiral forces in nuclear matter \cite{WhiteheadLH2019}, the \emph{ab initio} symmetry-adapted no-core shell model (SA-NCSM) \cite{LauneyMD_ARNPS21} for the astrophysically relevant low-energy regime based on the Green’s function method \cite{BurrowsL23}, and the microscopic structure-based methods \cite{BlanchonDAV2015,SargsyanPKE2023} using Feshbach projection \cite{Feshbach1958}.
 Furthermore, channel-dependent effective inter-cluster interactions can be constructed in the framework of the resonating group method (RGM) and no-core shell model with continuum (NCSMC) \cite{QuaglioniN09,BaroniNQ13}, as well as the symmetry-adapted SA-RGM applicable to the intermediate-mass region \cite{MercenneLDEQSD21}.

In this study, we focus on important ingredients for constructing such effective interactions that account for the microscopic structure of the reaction fragments (or clusters).
Specifically, using the $A$-body SA-NCSM description of the composite system with correct asymptotics at large distances, we study a nucleon plus target partitioning, which is important for studies of transfer, knockout and radiative capture reactions. 

In general, any nuclear many-body system can be described by single-nucleon spectroscopic overlaps. The norms of these overlaps are called spectroscopic factors (SFs). These overlaps can be derived as solutions of the Schr\"odinger equation with an effective inter-cluster potential, 
and thus convey important information about the interaction of a single nucleon with the target nucleus.
For example, the solutions of the  Schr\"odinger equation derived from the single-particle Green's function equation of motion \cite{CAPUZZI2000223,PhysRevC.66.034313} provide normalized spectroscopic overlaps, with SFs deduced from the energy derivative of the effective potential (see, e.g, Eq. (10) of \cite{PhysRevC.66.034313}). Equally, the spectroscopic overlaps with their SFs can be obtained by solving an inhomogeneous equation with a source term (see, e.g., \cite{BangGPV1985, PinkstonS1965}).
%

Historically, SFs have been used as a measure of single-nucleon clustering in nuclei. Experimentally, they are extracted from direct reaction measurements, as a normalization factor in the reaction cross section, which however takes into account any deviations from the model employed. Here, we provide SFs that are directly derived from many-body first-principle solutions, and as such, at infinite model spaces can be considered as a true measure of the single-nucleon clustering in nuclei. Indeed, various direct-reaction calculations of 
cross-sections 
are often reduced to using well-informed SFs and global optical potentials as inputs. 
In addition, unlike cross-sections that largely change depending on the energy of the projectile and the mass range of the target, SFs provide a simpler quantity for comparison of different systems. Even though SFs are model dependent, when extracted from different types of reactions they can be indicators of the physical relevance of the underlying model
 (see the comprehensive review \cite{AumannBBB2021} and references therein, as well as, e.g., \cite{KaySF2013}).
Furthermore, SFs can be directly used in constructions of optical potentials as they can be interpreted as a measure of coupling strengths for different reaction channels \cite{SargsyanPKE2023}. 


The asymptotic form of a spectroscopic overlap function is characterized by the asymptotic normalization coefficient (ANC). 
ANCs can be extracted from experimental data with fewer assumptions as compared to SFs, given reactions (e.g., well below the Coulomb barrier) that probe only the asymptotic part of the spectroscopic overlaps. Much of the work on ANCs has been motivated by their connection to astrophysical cross sections, however, they are also important tests of long-range nuclear physics.

In this paper we present 
calculations of spectroscopic overlap functions, SFs and ANCs using the \emph{ab initio} SA-NCSM for a series of Li isotopes. We compare the SFs and ANCs to the values deduced from experiments and Quantum Monte Carlo (QMC) calculations \cite{BridaPW2011, NollettW2011}. 
Specifically, the SA-NCSM \cite{LauneyDD16,DytrychLDRWRBB20,LangrDLD18,LangrDDLT19,1937-1632_2019_0_183} uses a physically relevant symmetry-adapted (SA) basis that can achieve significantly reduced  model spaces compared to the corresponding complete ultra-large model spaces, without compromising the accuracy of results for various observables \cite{DytrychHLDMVLO14,LauneyDD16,BakerLBND20}. The SA basis enables the SA-NCSM to accommodate contributions from more shells and to describe heavier nuclei, such as $^{20}$Ne \cite{DytrychLDRWRBB20}, $^{21}$Mg \cite{Ruotsalainen19}, $^{22}$Mg \cite{Henderson:2017dqc}, $^{28}$Mg \cite{WilliamsBCD2019},  as well as $^{32}$Ne and $^{48}$Ti \cite{LauneySOTANCP42018,LauneyMD_ARNPS21}.
Given the access of the SA-NCSM to these nuclear systems,
the methods developed in this work can be extended to the medium-mass region ($A\lesssim 50$) and nuclei with enhanced radii, deformation and clustering, 
especially those near the drip lines.

\section{Theoretical framework}
The \emph{ab initio} SA-NCSM is a no-core shell model that uses an  \SpR{3}-coupled or \SU{3}-coupled basis, referred to as ``symmetry-adapted" (see Refs. \cite{LauneyDD16,DytrychLDRWRBB20} and the references therein). It builds upon a harmonic oscillator (HO) single-particle basis, similar to the NCSM \cite{NavratilVB00,NavratilVB00b},
where the HO major shells are separated by a parameter
$\hbar \Omega$. 
The model space is limited by an \Nmax~cutoff which is the largest number of total excitation quanta considered above the lowest HO configuration for a given nucleus. The nuclear Hamiltonian utilized in the SA-NCSM is non-relativistic and uses translationally invariant nucleon-nucleon plus Coulomb interactions. Since we perform calculations in laboratory coordinates, we eliminate the spurious center-of-mass excitation states from the low-lying spectrum with a Lawson term \cite{Lawson74, DytrychMLDVCLCS16}. The Lawson procedure uses a Lagrange multiplier term that is added to a Hamiltonian expressed in laboratory-frame coordinates, $H+\lambda N_{\rm CM}$ where $N_{\rm CM}$ is the operator that counts the number of CM excitations and $n_{\rm CM}$ is its eigenvalue. For a typical value of $\lambda \sim 50$ MeV, the nuclear states of interest (with energy  $\leq 30$ MeV) have wave functions that are free of center-of-mass excitations ($n_{\rm CM}=0$), while CM-spurious states ($n_{\rm CM} > 0$) lie much higher in energy. For a given nucleus the SA-NCSM constructs the nuclear interaction Hamiltonian and calculates its eigenvalues and eigenvectors. The eigenvectors are subsequently used for calculations of the nuclear observables. As the model space increases the calculations approach the exact value. The results become independent of the HO parameter $\hbar \Omega$ at the \Nmax $\rightarrow \infty$ limit or at convergence.
The SA-NCSM results exactly match those of the NCSM for the same interaction within a given complete \Nmax~model space. The use of symmetries in SA-NCSM allows one to select the model space by considering only the physically relevant subspace, which is only a fraction of the corresponding  complete \Nmax~space. 
The calculations throughout this paper are performed using the \NNLOopt~chiral interaction \cite{Ekstrom13} that are not  renormalized (e.g., by using the SRG technique) in the nuclear medium. This interaction minimizes the effect of the three body forces and has been shown to give a good description of nuclear structure and reaction observables (see, \eg, Refs. \cite{Ruotsalainen19,WilliamsBCD2019, BurrowsEWLMNP19,BakerLBND20,Miller22}).


To calculate the spectroscopic overlap for a single-nucleon projectile ($a=1$) and a composite nucleus of mass $A$, we use the SA-NCSM, which provides wave functions for the $A-1$ and $A$ systems in laboratory coordinates (L) with the center-of-mass contribution exactly factored out and in the lowest HO state.
The spectroscopic overlap has the form \cite{Navratil2004,NavratilBC2006}:

\begin{widetext}
\begin{eqnarray}
 u_{A-1 \alpha_1 I_1; l {1\over 2}j}^{A\alpha J}(r) &=& \sum_n  R_{n l}(r)\braket{\Phi_{A\alpha J}}{\mathcal{A}\Phi_{A-1 \alpha_1 I_1; l {1\over 2}j}^J} \nonumber \\
 &=& \sum_n  R_{n l}(r)\frac{1}{\braket{nl00l}{00nll}_{1/A-1}}\frac{1
 }{\Pi_J} 
 (-1)^{\frac{n-l}{2}}\RedME{A \alpha J}{\ad_{(n\,0)lj}}{A-1\alpha_1 I_1}_{\mathrm{su3;L} },
 \label{eq:overlapj}
\end{eqnarray}
\end{widetext}
where the antisymmetrization $\mathcal{A}$ between the two clusters is included, as shown in Eq. (16) in Ref. \cite{Navratil2004}, and where the matrix element, $\RedME{}{}{}$, is reduced with respect to angular momentum, but calculated in the \SU{3} basis using a creation operator expressed as an \SU{3} tensor $\ad_{(n \, 0) l j} \equiv \ad_{n l j}$ 
HO shell number $n$ (cf. \cite{NavratilBC2006} for the conventional shell-model notations in terms of the HO radial quantum number $n_r=\frac{n-l}{2}$),
 $I_1$ and $J$ are the total angular momenta of the target and composite nuclei, respectively, and $\Pi_J=\sqrt{2J+1}$. The coupling of the orbital momentum $l$ of the nucleon with its spin (1/2) yields $j$ (we note that we work in a proton-neutron formalism and isospin is not a good quantum number of the basis; also, we omit the nucleon spin label 1/2 from the formulae below).
 The labels $\alpha_1$ and $\alpha$ denote the additional quantum numbers needed to characterize the eigenstates. 
 The $R_{nl}(r)$ is the radial wavefunction that is positive at origin.
 In the SA-NCSM the reduced matrix element is calculated using the eigenvectors of the initial and final many-body states expressed in the SU(3)-coupled basis. 
 The bra and ket eigenstates correspond to the laboratory-frame wavefunctions of the composite and target nuclei, respectively, calculated from the many-body theory. A Lawson procedure ensures that both eigenfunctions can be factorized to an intrinsic wavefunction  and a center-of-mass (c.m.) wavefunction that is in the lowest HO state ($n_{\rm c.m.}=0$ and $l_{\rm c.m.}=0$) \cite{Lawson74}. The Talmi-Moshinsky bracket that transforms the  c.m.
 coordinates of the two clusters in the laboratory frame to a relative distance $r$ between the clusters 
for translationally invariant spectroscopic overlaps is given by \cite{Navratil2004}:
 \begin{equation}
 \braket{nl00l}{00nll}_{a/A-a} = (-1)^l \Big(\frac{A-a}{A} \Big)^{n/2},
 \end{equation}
where $a$ and $A-a$ are the numbers of nucleons in the clusters.
 
The norm of the spectroscopic overlap,
\begin{eqnarray}
 S_{A-1 \alpha_1 I_1; lj}^{A\alpha J} &=& \int_0^\infty |u_{A-1 \alpha_1 I_1; lj}^{A\alpha J}(r)|^2 r^2 dr\\
 &=& \sum_n |u_{\nu lj;n}^{J}|^2 
\label{eq:SF}
\end{eqnarray}
is called the spectroscopic factor (SF), where in the second line of Eq. (\ref{eq:SF}) we use the channel notation $\nu=\{A\alpha,A-1 \alpha_1 I_1\}$, along with $u_{\nu lj}^{J}(r)=\sum_n R_{nl}(r) u_{\nu lj;n}^{J}$.

For a nucleus of $A$ particles, partitioned into two clusters $A-a$ and $a$, the cluster wavefunction is considered in two regions: interior, where the wavefunction is driven by the inter-nucleon interactions and  is given by Eq. (\ref{eq:overlapj}) in this study, and exterior, where the only interaction between the clusters is the Coulomb force and the exact Coulomb eigenfunctions are used.

For bound states the exterior wavefunction for two clusters with relative angular momentum $l$ and separated at distance $r$ is given by the asymptotically decaying Whittaker function \cite{DreyfussLESBDD20, ThompsonN2009, DescouvemontB10}:
\begin{equation}
    W_{-\eta, l+\frac{1}{2}}(2kr) \xrightarrow[r \to \infty]{} (2kr)^{-\eta}e^{-kr},
\end{equation}
with $k=\sqrt{2\mu B}/\hbar$, where $B$ is the cluster separation energy and $\mu$ is the reduced mass of the two clusters $A-a$ and $a$, and $\eta=Z_a Z_{A-a} \mu e^2/\hbar^2 k$ is the Sommerfeld parameter. 
The amplitude of the exterior wavefunction at large distances $r$ is called asymptotic normalization coefficient (ANC), hence the exterior bound state wavefunction is given as 
\begin{equation}
  \phi_{\nu lj}^{J, \, \mathrm{ext}}(r) \approx C_{\nu lj}^J \frac{W_{-\eta, l+\frac{1}{2}}(2kr)}{r}.
  \label{eq:ext_wfn}
\end{equation}
Here, $C_{\nu lj}^J$ corresponds to the ANC, where  $\nu$ represents all quantum numbers needed to fully characterize the respective states of the two clusters, and contains their parities and total angular momenta. 

Asymptotically, for large $r$ the interior cluster wavefunction (or the spectroscopic overlap) $u_{\nu lj}^{J}(r)$ should approach the exterior wavefunction in Eq. (\ref{eq:ext_wfn}). To extract the ANCs, one can match the interior wavefunction to the exterior one at the channel radius $r_c$ between the centers of masses of the two clusters similar to Ref. \cite{BridaPW2011}:  
\begin{equation}
  C_{\nu lj}^J \approx \frac{r_c u_{\nu lj}^{J}(r_c)}{ W_{-\eta, l+\frac{1}{2}}(2kr_c)}.
  \label{eq:ANC}
\end{equation}
This formula assumes that the interior wavefunctions at the channel radius is approximately equal to the long-range Coulomb solution.  Other methods for calculating ANCs are described in Refs. \cite{Brune_PRC66_2002,NollettW2011, Timofeyuk2010,DreyfussLESBDD20}. 
Since matching the tail of the overlap function may change  the SF, we modify Eq. (\ref{eq:ANC}) to preserve the SF of Eq. (\ref{eq:SF}):
\begin{widetext}
\begin{equation}
  C_{\nu lj}^J \approx \sqrt{ S_{A-1 \alpha_1 I_1; lj}^{A\alpha J}} \Bigg[\Big(\frac{ W_{-\eta, l+\frac{1}{2}}(2kr_c)}{r_c u_{\nu lj}^{J}(r_c)}\Big)^2  \int_0^{r_c} (r u_{\nu lj}^{J}(r) )^2dr + \int_{r_c}^\infty \big( W_{-\eta, l+\frac{1}{2}}(2kr) \big)^2 dr \Bigg]^{-1/2}.
  \label{eq:ANC_const_SF}
\end{equation}
\end{widetext}
Clearly, for sufficiently large $r_c$, the last equation reduces to Eq. (\ref{eq:ANC}).
As often done, we choose the channel radius to maximize the ANC, or equivalently, to match the logarithmic derivatives. In this study, the channel radii are found to be typically large, so Eqs. (\ref{eq:ANC_const_SF}) and (\ref{eq:ANC}) yield practically the same outcome. 



\section{Results and discussions}

\begin{figure}[ht]
    \centering
    \includegraphics[width=0.49\textwidth]{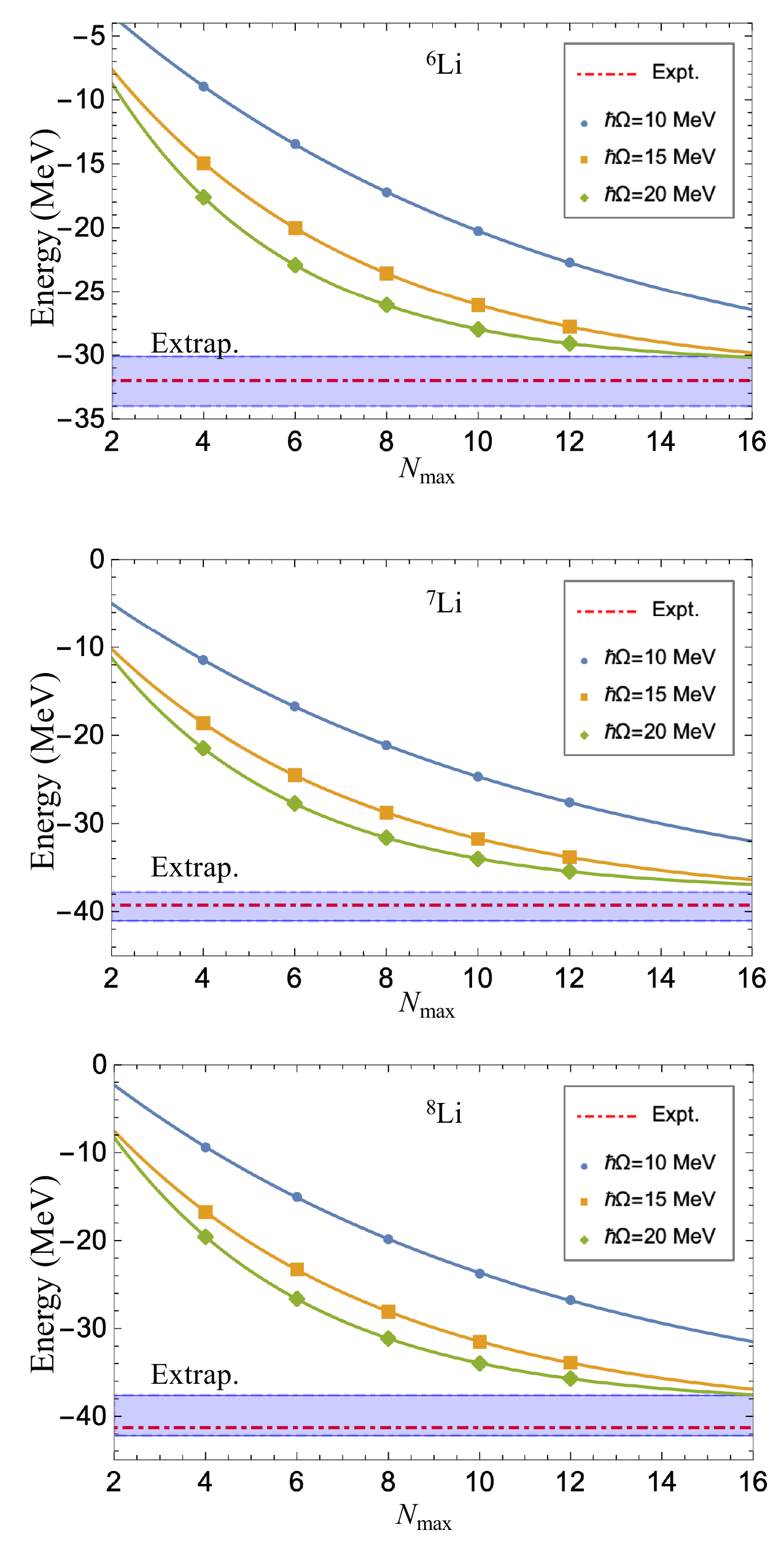}
    \caption{Ground state energies of \Li{6}, \Li{7} and \Li{8} from SA-NCSM calculations with \NNLOopt~interaction with HO parameters \hw=10, 15 and 20 MeV, and compared to the experiment. The blue bands indicate the extrapolated energies (denoted as ``Extrap.''). The uncertainties of the extrapolated values are from fitting and \hw~variance.}
   \label{fig:6Li_8Li_E}
\end{figure}

\begin{figure}[h]
    \centering
    \includegraphics[width=0.49\textwidth]{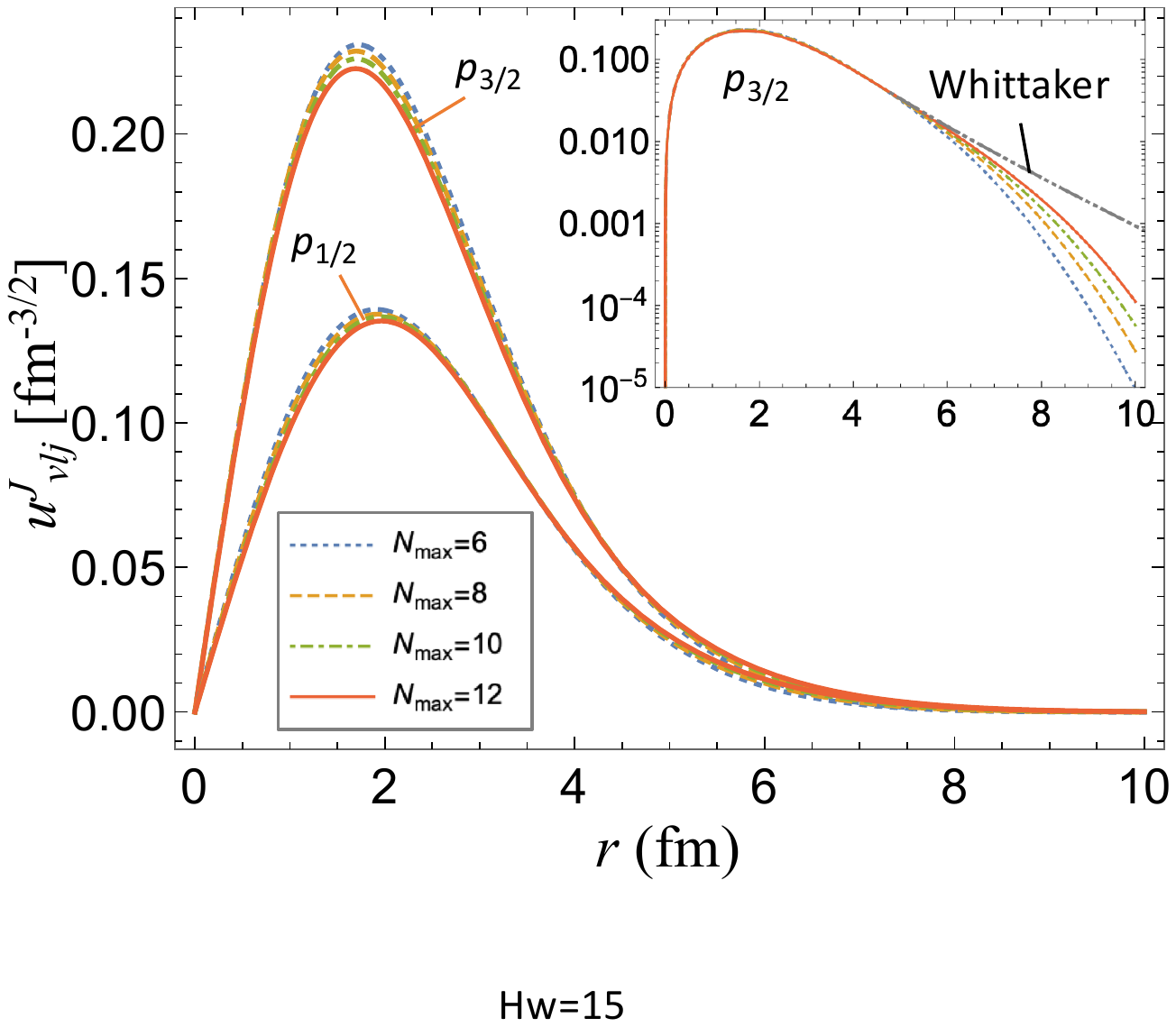}
    \caption{Single-nucleon overlaps $\braket{^7\mathrm{Li}}{^6\mathrm{Li}+{\rm n}}$ in \Nmax=6 to 12 with \hw=15 MeV for partial waves $p_{1/2}$ and $p_{3/2}$ vs. the separation between $^6\mathrm{Li}$ and the neutron. Inset: same, but for $p_{3/2}$ only and in $\log_{10}$ scale on the ordinate axis along with the exterior Whittaker function.}
   \label{fig:6Li_ov_Nmax}
\end{figure}


We present SA-NCSM calculations of single-nucleon overlaps $\braket{^7\mathrm{Li}}{^6\mathrm{Li}+{\rm n}}$, $\braket{^8\mathrm{Li}}{^7\mathrm{Li}+{\rm n}}$ and $\braket{^9\mathrm{Li}}{^8\mathrm{Li}+{\rm n}}$ for the ground states of the Li isotopes. 
We match the calculated overlaps to the exterior Whittaker function to calculate the ANCs and compare them to the experimentally deduced values. 
Using the wavefunctions from the many-body calculations and Eq. (\ref{eq:overlapj}) we calculate the single-nucleon overlaps for partial waves $p_{1/2}$ ($l=1, j=1/2$) and $p_{3/2}$ ($l=1, j=3/2$).

The corresponding SA-NCSM ground-state energies of the Li isotopes  are found to be on a converging trend in sufficiently large model spaces (Fig. \ref{fig:6Li_8Li_E}). Since the calculations are performed in a finite model space, the ground state energies converge to the infinite-space results from above. We perform calculations using the HO parameter  \hw=10, 15 and 20 MeV and for each of the \hw~values, we extrapolate to the infinite space using a three-parameter exponential formula, similar to Ref. \cite{MarisVS09}:
\begin{equation}
E(N_\mathrm{max})=E(\infty)+a \, \mathrm{exp}(-c N_\mathrm{max}),
\label{eq:extrap}
\end{equation}
where $E(\infty)$ is the energy at the infinite model space. The error bars indicate the combined uncertainty due to the \hw~variance and the fitting. The extrapolated values are in a good agreement with the experiment (Fig. \ref{fig:6Li_8Li_E}).


\subsection{Spectroscopic factors}
\begin{figure*}[btp]
    \centering
    \includegraphics[width=0.99\textwidth]{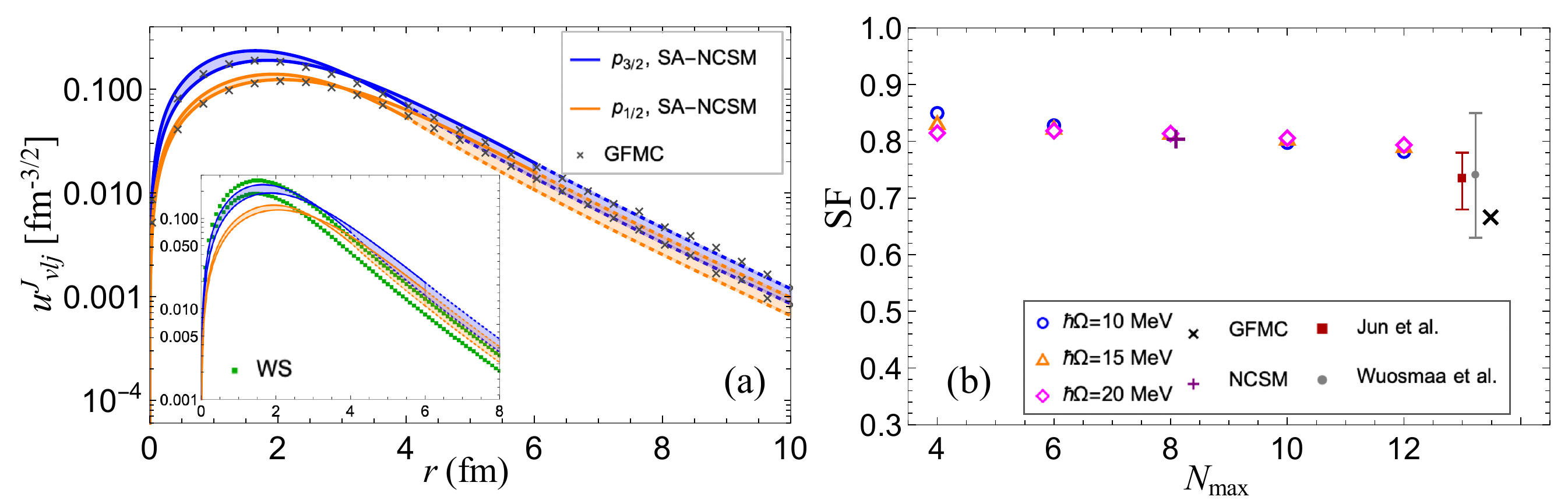}
    \caption[Single-nucleon overlaps and spectroscopic factors of $\braket{^7\mathrm{Li}}{^6\mathrm{Li}+{\rm n}}$]{(a) Single-nucleon overlaps of the $^7$Li ground state with the $^6\mathrm{Li}+{\rm n}$ in \Nmax=12 for partial waves $p_{1/2}$ and $p_{3/2}$ vs. the separation between $^6\mathrm{Li}$ and the neutron, $r$, compared to GFMC results from Ref. \cite{QMC_database}. The dotted lines correspond to the exterior Whittaker function. The shaded bands indicate the uncertainty due to the \hw~variance from 10 to 20 MeV (interior to exterior wavefunction matching radii differ for different \hw, see text for details). For all but the first two points, the GFMC uncertainties are smaller than the marker size on the plot. Inset: SA-NCSM overlaps compared to a typical Woods-Saxon parameterization (see text for details). (b) Calculated SFs with the increasing model space \Nmax~and compared to the experimentally deduced values from Jun \etal~\cite{JunZBX2010} and Wuosmaa \etal~ \cite{WuosmaaSRGH2008}. Also shown are SFs calculated in the NCSM from Ref. \cite{Navratil2004} and GFMC from Ref. \cite{BridaPW2011}. 
    }
   \label{fig:6Li_7Li_Overlap}
\end{figure*}

The ground states of Li isotopes that differ in mass by one nucleon have opposite parities. Since the parity of the nucleon is given by $(-1)^l$, $l$ must be odd to preserve the parity. In addition, as it was shown in Ref. \cite{NollettW2011} and observed in our calculations, the $l=3$ overlaps are much smaller than the $l=1$ overlaps (see Table \ref{tab:SF}). Hence, we limit our analysis to  only $l=1$, $p_{1/2}$ ($j=1-1/2$) and $p_{3/2}$ ($j=1+1/2$) partial waves. For these two values of $j$ we explore the dependence of the $\braket{^7\mathrm{Li}}{^6\mathrm{Li}+{\rm n}}$ overlaps on the model space size for a fixed \hw~(Fig. \ref{fig:6Li_ov_Nmax}). With increasing \Nmax~model space, the changes between successive curves become smaller indicating convergence of the overlaps. Similar dependence is observed for $\braket{^8\mathrm{Li}}{^7\mathrm{Li}+{\rm n}}$ and $\braket{^9\mathrm{Li}}{^8\mathrm{Li}+{\rm n}}$ overlaps with a fixed \hw. 
It should be noted that peripheral reactions and long-range observable, such as rms radii and quadrupole moments, are sensitive to the overlap at large distance (tail). Indeed, the tail approaches the exact Whittaker functions  with increasing model space size (Fig. \ref{fig:6Li_ov_Nmax}, inset). For example, in the case of ${^6\mathrm{Li}+{\rm n}}$, at $N_\mathrm{max}=12$ the tail coincides with the exact solution up to about 7 fm, thereby allowing for matching. 
Since the overlaps decay quickly at larger radii, it is more informative to present them in a logarithmic scale (Fig. \ref{fig:6Li_7Li_Overlap}a). In this figure, we use the exact Whittaker function 
at large distances. Furthermore, the overlaps are represented as bands of values due to the variance of the HO parameter \hw~ from 10 to 20 MeV, typical for nuclei in this mass range. The matching radii are different depending on the HO parameter. For comparison, overlaps from \emph{ab initio} GFMC (gray crosses) as well as a typical Woods-Saxon (WS) potential (Fig. \ref{fig:6Li_7Li_Overlap} inset) are presented. The depth of the WS potential has been fitted to reproduce the experimental neutron separation energy of $^7$Li for each partial wave: $V_0=-71.95$ MeV for $p_{1/2}$ and $V_0=-61.31$ MeV for $p_{3/2}$, along with $R_0=1.25A^{1/3}$ fm radius and  $a=0.65$ diffuseness, and a spin-orbit term with $V_\mathrm{SO}=6$ MeV depth and the same $R_0$ and $a$. The overlaps from the WS solutions have been normalized to reproduce the same SFs as the SA-NCSM ones. Even though SA-NCSM and GFMC use different NN interactions, both \textit{ab initio} approaches yield very similar overlaps. In contrast, the WS overlaps peak at higher values for both partial waves and are below the \emph{ab initio} overlaps at long distances, which would result in smaller ANCs. We integrate the overlaps to obtain the SFs for each of the partial waves using Eq. (\ref{eq:SF}), with the total SF given by the sum of the SF for both partial waves. For $\braket{^7\mathrm{Li}}{^6\mathrm{Li}+{\rm n}}$ the calculated SFs converge towards the experimentally deduced value from Ref. \cite{JunZBX2010} as the model space increases (Fig. \ref{fig:6Li_7Li_Overlap}b). In addition, we compare our calculations to the values from GFMC \cite{BridaPW2011} and NCSM \cite{Navratil2004} (also, cf. Ref. \cite{FernandezMPM2023}).

\begin{figure*}[t]
    \centering
    \includegraphics[width=0.99\textwidth]{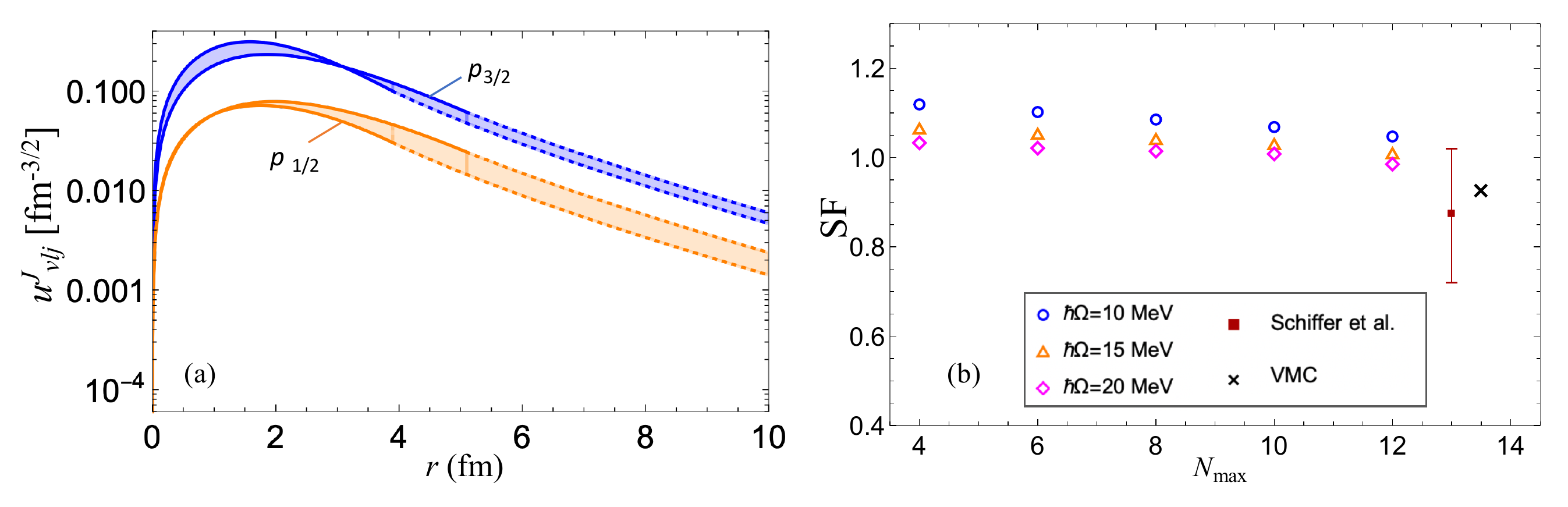}
    \caption{Same as Fig. \ref{fig:6Li_7Li_Overlap} but for $^7\mathrm{Li} + n$. The experimentally deduced SF is from Schiffer \etal ~\cite{Schiffer8Li1967}, and the VMC result is from Ref. \cite{QMC_database}. }
   \label{fig:7Li_8Li_Overlap}
\end{figure*}

\begin{figure*}[t]
    \centering
    \includegraphics[width=0.99\textwidth]{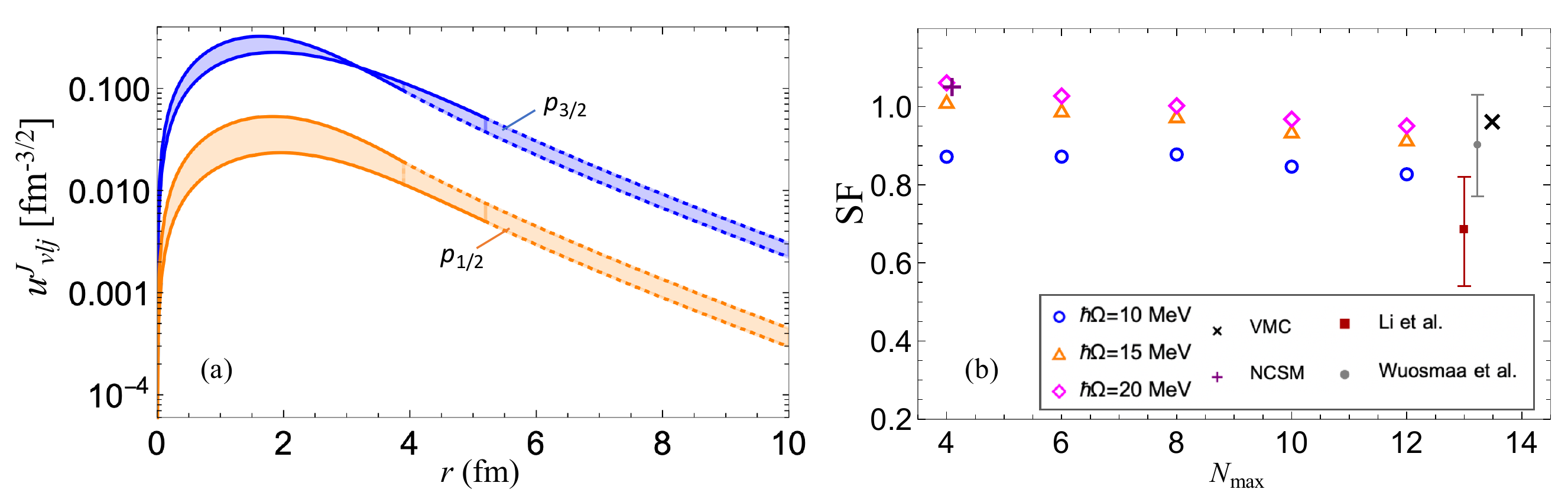}
    \caption{
    Same as Fig. \ref{fig:6Li_7Li_Overlap} but for $^8\mathrm{Li} + n$. The experimentally deduced SFs are from Li \etal~ \cite{Li9Li2005} and Wuosmaa \etal~ \cite{Wuosmaa9Li2005}. The values from the VMC and NCSM are from Refs. \cite{QMC_database} and \cite{Navratil2004}, respectively. }
   \label{fig:8Li_9Li_Overlap}
\end{figure*}

\begin{figure}[t]
    \centering
    \includegraphics[width=0.49\textwidth]{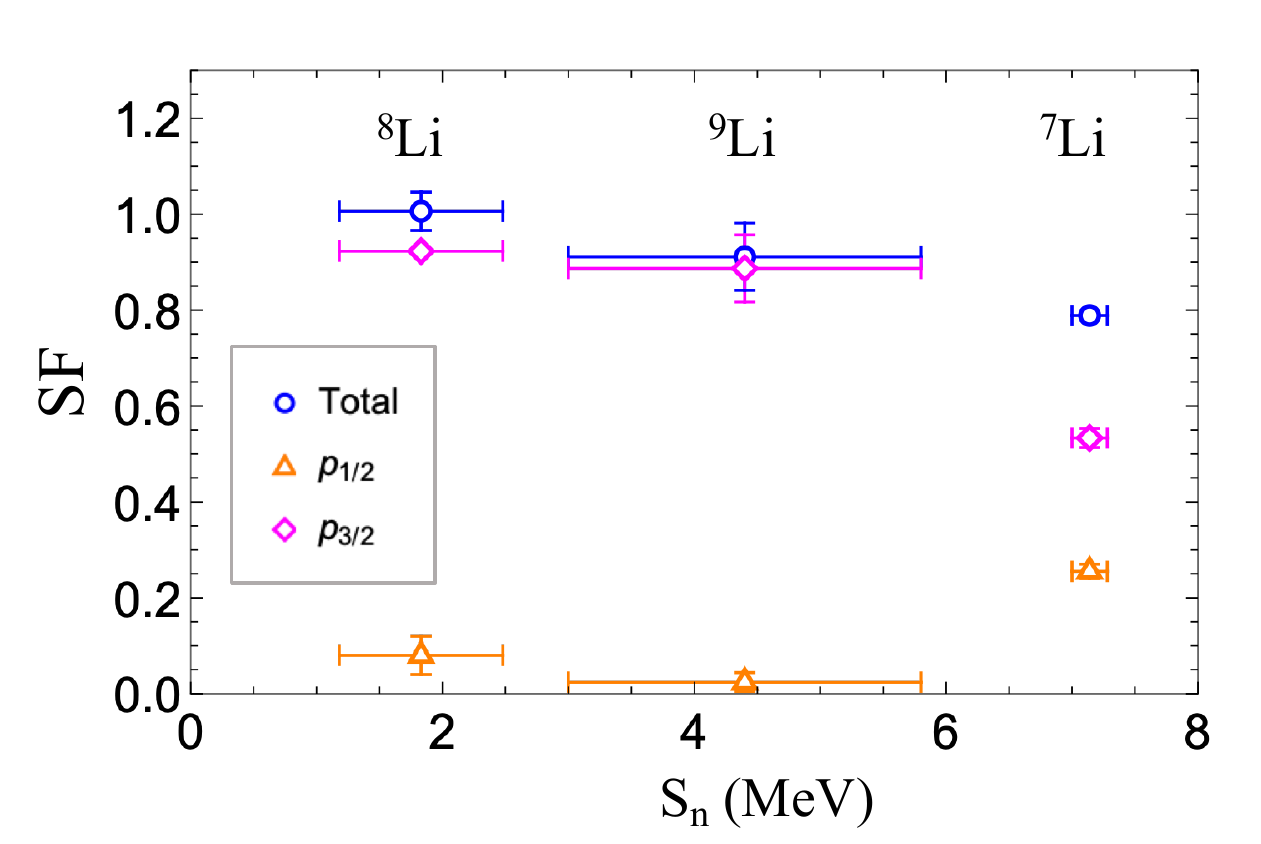}
    \caption{ Calculated SFs in \Nmax=12 vs the extrapolated neutron separation energies ($S_n$) from calculations. Uncertainties on $S_n$ are from \hw~variance and SA model space selection.}
   \label{fig:SFvsSn}
\end{figure}

A similar behavior is found for the $\braket{^8\mathrm{Li}}{^7\mathrm{Li}+{\rm n}}$ and $\braket{^9\mathrm{Li}}{^8\mathrm{Li}+{\rm n}}$ overlaps (Figs. \ref{fig:7Li_8Li_Overlap} and \ref{fig:8Li_9Li_Overlap}). In $\braket{^8\mathrm{Li}}{^7\mathrm{Li}+{\rm n}}$ the spread of values due to the \hw~variance is small in the short-range part of the wavefunction but becomes more apparent at larger radii. This shows that the description of the long-range part of the wavefunction is sensitive to the \hw~values for a given \Nmax~model space.
The larger spread in the $p_{1/2}$ overlap of $\braket{^9\mathrm{Li}}{^8\mathrm{Li}+{\rm n}}$ is due to its smaller values (by an order of magnitude) compared to the $p_{3/2}$ (Fig. \ref{fig:8Li_9Li_Overlap}a). 
For both $\braket{^8\mathrm{Li}}{^7\mathrm{Li}+{\rm n}}$ and $\braket{^9\mathrm{Li}}{^8\mathrm{Li}+{\rm n}}$ the calculations of SFs with all HO frequencies considered are converging to the uncertainty range of the experimentally-deduced results (Figs. \ref{fig:7Li_8Li_Overlap}b and \ref{fig:8Li_9Li_Overlap}b). A particularly interesting case is $^8$Li+n, where the two experimentally deduced SFs reported in Fig. \ref{fig:8Li_9Li_Overlap}b agree with each other within the uncertainties, however, the calculated SFs are closer to 
the outcome of Wuosmaa \etal~(Ref. \cite{Wuosmaa9Li2005}), as shown in the figure. Since both experiments use the same reaction, $^{2}$H($^{8}$Li,p)$^{9}$Li, at a similar energy regime but different data sets, this suggests that the data analysis in Ref. \cite{Li9Li2005} has likely underestimated the neutron $p$-wave channel contribution to $^9$Li. Nonetheless, we note that all these outcomes are considered in a good agreement.
For more experimental evaluations of $\braket{^9\mathrm{Li}}{^8\mathrm{Li}+{\rm n}}$ SFs see Ref. \cite{KanungoABD2008}.

Among the three cases we present, the calculated SFs in the largest model spaces are close to one, except for the case of $^6$Li+n. This suggests a more complicated structure of the $^7$Li ground state, which could be related to a low-lying $\alpha +t$ threshold that lies closer to the ground state compared to the $^6$Li+n threshold. In comparison, the neutron channel is the lowest in energy for both $^8$Li and $^9$Li. It is therefore interesting to study the $\alpha$ overlap for $^7$Li and the effects of alpha clustering. Further examining dependence of the SFs on the neutron threshold, we find a slow decrease of SFs as the separation energy increases (Fig. \ref{fig:SFvsSn}), similar to Fig. 18 in Ref. \cite{AumannBBB2021}. This suggests a stronger single-particle clustering when the neutron binding is weaker, although we leave a detailed analysis for future SA-NCSM studies that will span a broader region of nuclei. We note that the neutron thresholds reported in Fig. \ref{fig:SFvsSn} are determined from  \Nmax $\rightarrow \infty$ extrapolations of the binding energies, with uncertainties that take into account the model space selection and \hw~ variance. All the thresholds are in agreement with the experimental values. 



\subsection{Spectroscopic factors and amount of clustering}
The spectroscopic overlaps are calculated in this paper using Eq. (\ref{eq:overlapj}) based on the $\ket{\mathcal{A}\Phi_{A-1 \alpha_1 I_1; l {1\over 2}j}^J}$ two-cluster states. 
%
%
The antisymmetrization guarantees that the Pauli exclusion principle is correctly taken in account within the ${ A }$-nucleon system, however, it renders the set of cluster wave functions to be neither normalized nor orthogonal.
This non-orthonormality prevents one to interpret ${ { | { u }_{ A-1 { \alpha }_{ 1 } { I }_{ 1 } ; l j }^{ A \alpha J } (r) | }^{ 2 } }$ in Eq. (\ref{eq:SF}) as a probability on an absolute scale (e.g., see \cite{LovasLIVD1998,Rodkin2020}), and to relate the spectroscopic factors  to the amount of clustering.
Therefore, in order to probe the amount of clustering, it is necessary to utilize orthonormalized cluster wave function 
\cite{LovasLIVD1998,Rodkin2020}:

\begin{equation}
 \mathcal{S}_{\nu l}^{J} =  \sum_{j_1 n_1 ,j_2 n_2,k} u_{\nu lj_1;n_1}^J e^{(k)}_{\nu lj_1;n_1} {1 \over N_{\nu l; k}} e^{(k)}_{\nu lj_2;n_2} u_{\nu lj_2;n_2}^J,
\label{eq:SFn}   
\end{equation}
where $N_{\nu l;k}$ is the $k$th eigenvalue of the norm matrix $\mathcal{N}_{\nu l}$, and $e^{(k)}_{\nu lj_1;n_1}$ are the components of the corresponding eigenstate. While, in general, the norm matrix mixes orbital momenta, in this study we neglect $l=3$, as discussed above. We also note that SFs are calculated before matching and remain practically unchanged after matching, which justifies the use of configuration representation in Eq. (\ref{eq:SFn}).
The orthonormalization process involves the inversion of the norm kernel, which in configuration representation is a matrix whose elements are given by the overlap between two cluster wave functions (see, e.g., \cite{QuaglioniN09}). 
Typically, the amount and magnitude of the off-diagonal elements reflect the non-orthogonality of the cluster wave functions under consideration. 

To test the amount of clustering $\mathcal{S}_{\nu l}^{J}$, in  this study, the norm kernel is computed and studied through the use of the symmetry-adapted RGM \cite{MercenneLDEQSD21}. We find that the off-diagonal elements involving the nucleon projectile being in ${ p_{3/2} }$ or ${ p_{1/2} }$ are extremely small (around $10^{-3}$). Hence their contribution to $\mathcal{S}_{\nu l}^{J}$
is expected to be negligible.
As a consequence, only diagonal elements were considered in the computation 
with
$\mathcal{S}_{\nu, l=1}^J=\mathcal{S}_{\nu p_{1/2}}^J+\mathcal{S}_{\nu p_{3/2}}^J$. Indeed, for $^6$Li+n the Hilbert-Schmidt norm of $\mathcal{N}_{\nu l}-\mathbb{1}$ is only $0.01$, whereas the relative difference of the spectroscopic factors given in Fig. \ref{fig:6Li_7Li_Overlap}b compared to the one using normalized cluster wave functions is only $1.6\%$-$2.1\%$ across \hw=10-25  MeV. 

To summarize, for the systems under considerations,
the norm matrix is approximately the identity matrix, yielding SFs that are practically the same as the measure for clustering $\mathcal{S}$. We note that, in general, the effect of the norm matrix should not be neglected, especially when more partial waves need to be considered.

\subsection{Asymptotic normalization coefficients}
Obtaining ANCs directly from matching the overlaps to Eq. (\ref{eq:ANC_const_SF}) can be challenging for the many-body methods that use HO basis, since the asymptotics at large radii is affected by the model space cutoff. To ensure the correct asymtotics, large model spaces are required, as illustrated in the inset of Fig. \ref{fig:6Li_ov_Nmax}a. Alternative methods have been also developed to address this issue, e.g., see Ref. \cite{NollettW2011,Timofeyuk2010, Brune_PRC66_2002}. The challenges associated with each of the methods are reviewed in Ref. \cite{timofeyuk2014overlap}.
In addition, the extractions of ANCs from Eq. (\ref{eq:ANC_const_SF}) requires the separation energy $B$. To be fully consistent in determining the ANCs from overlaps, one should use theoretically calculated separation energies $B_\mathrm{th}$. Nevertheless, in most models the experimental value $B_\mathrm{exp}$ is used to make the ANCs practical for reaction calculations, since even small deviations of $B_\mathrm{th}$ from $B_\mathrm{exp}$ can affect the ANCs \cite{NollettW2011}. 

\begin{figure*}[]
    \centering
    \includegraphics[width=0.99\textwidth]{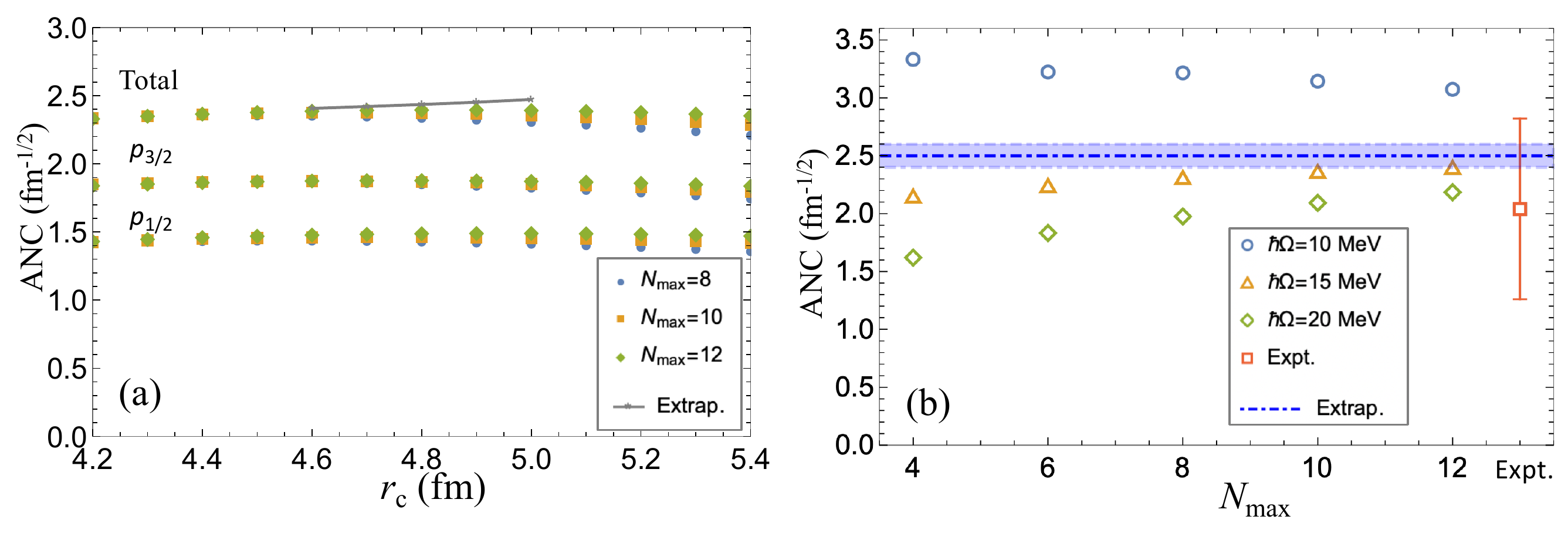}
    \caption{ANCs for the $\braket{^7\mathrm{Li}}{^6\mathrm{Li}+{\rm n}}$ (a)  as a functions of the channel radius $r_c$ with \hw=15 MeV, with the connected gray points showing the extrapolated results (denoted as ``Extrap.''), and (b) for \hw=10-20 MeV vs. the model space size and compared to the experiment \cite{GulamovMN1995} (denoted as ``Expt.''). The label ``Total'' in (a) denotes $\sqrt{\sum C_{lj}^2}$ for the $p_{1/2}$ and $p_{3/2}$ partial waves. The horizontal dot-dashed line with the band in (b) shows the extrapolation from \hw=15 and 20 MeV calculations with the uncertainty from \hw~variance. 
   }
   \label{fig:6Li_7Li_ANC}
\end{figure*}
 
 We calculate the ANCs by directly matching the spectroscopic overlap to the exterior Whittaker function and preserving the SF of the unmatched overlap. We choose a channel radius $r_c$ that maximizes the ANC (or equally, ensures slope continuity), and we use $B_{\rm exp}$, which falls within the calculated extrapolated energy for all cases under consideration, as shown in Fig. \ref{fig:6Li_8Li_E}.
 To report a parameter-free ANC (Table \ref{tab:ANC}), we use the Shanks transformation \cite{Shanks1955,Schmidt1941} for the \Nmax=8, 10 and 12 calculations, with \hw=15 and 20 MeV that are close to convergence with \Nmax; 
in addition, 
 the final estimate is required to be 
 independent from the channel radius, as illustrated in Fig. \ref{fig:6Li_7Li_ANC}a. 
 For the Li isotopes discussed here the fastest convergence of ANCs is observed for \hw=15 MeV, which appears to be the optimal \hw~value, that is, where the convergence of results is achieved at comparatively smaller model spaces, while other \hw~values demand larger \Nmax~to produce the same estimate. In particular, the  ANC for \hw=15 MeV for  $\braket{^7\mathrm{Li}}{^6\mathrm{Li}+{\rm n}}$ flattens around $r_c=4.6$-5 fm (Fig. \ref{fig:6Li_7Li_ANC}a), where the SA-NCSM overlap function indeed coincides with the Whittaker function, as shown in Fig. \ref{fig:6Li_ov_Nmax}.
 The total ANC for the $p_{1/2}$ and $p_{3/2}$ partial waves is calculated by $C_l=\sqrt{C_{p1/2}^2+C_{p3/2}^2}$ (using a diagonal norm matrix, which is an excellent approximation, as discussed above). Our prediction based on the extrapolated results of \hw=15 MeV and 20 MeV  is $2.5\pm0.1$ fm$^{-1/2}$ (cf. Table \ref{tab:ANC}), which is within the experimentally deduced range of $1.26 - 2.82$ fm$^{-1/2}$ \cite{GulamovMN1995}. Indeed, the \hw=15 and 20 MeV calculations are on converging trend with \Nmax, and at \Nmax=12 agree with the extrapolated estimates and with the experimentally deduced range (Fig. \ref{fig:6Li_7Li_ANC}b).

\begin{table}[tbh]
\begin{center}
\caption{\label{tab:SF} SA-NCSM calculations of SFs for \Nmax=12 model space.  }
\begin{tabular}{c c| c c c  } \hline \hline
Transition &  & $\hbar\Omega$=10 MeV & $\hbar\Omega$=15 MeV & $\hbar\Omega$=20 MeV \\ \hline
$\braket{^7\mathrm{Li}}{^6\mathrm{Li}+{\rm n}}$ &$p_{3/2}$ &0.51 &0.53 &0.54 \\
 &$p_{1/2}$ &0.27 &0.26 &0.25 \\
  &$f_{5/2}$ &   &$6.7 \times 10^{-5}$ &  \\
$\braket{^8\mathrm{Li}}{^7\mathrm{Li}+{\rm n}}$  &$p_{3/2}$ &$0.92$ &$0.92$ &  0.92\\
  &$p_{1/2}$ &0.12 &0.08 & 0.07 \\
  &$f_{5/2}$ & &$3.4 \times 10^{-3}$ &   \\
   &$f_{7/2}$ & &$1.1\times 10^{-4}$ &   \\
$\braket{^9\mathrm{Li}}{^8\mathrm{Li}+{\rm n}}$  &$p_{3/2}$ &$0.82$ & $0.89$ & 0.92 \\
  &$p_{1/2}$ &0.01 &0.02 & 0.03 \\
   &$f_{5/2}$ &  & $1.5 \times 10^{-3}$ &  \\
    &$f_{5/2}$ &  & $1.9 \times 10^{-6}$ &   \\
\hline\hline
\end{tabular}
\end{center}
\end{table}

\begin{figure*}[ht]
    \centering
    \includegraphics[width=0.99\textwidth]{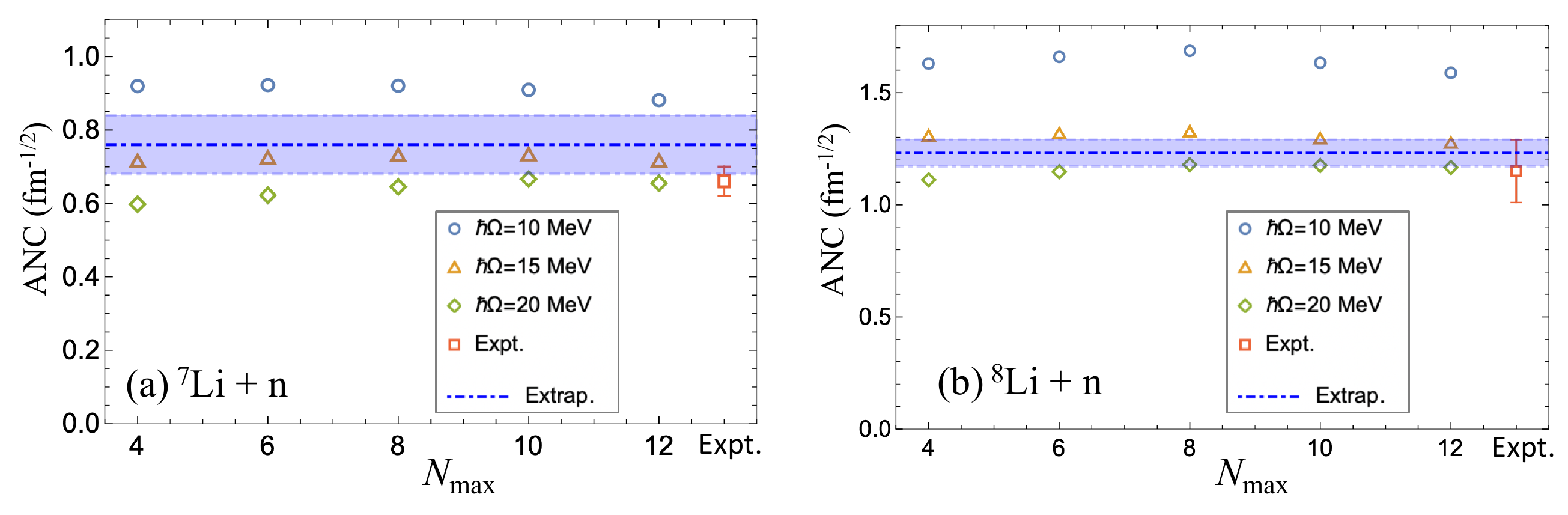}
    \caption{Total ANCs for the (a)$\braket{^8\mathrm{Li}}{^7\mathrm{Li}+{\rm n}}$ and (b) $\braket{^9\mathrm{Li}}{^8\mathrm{Li}+{\rm n}}$ for \hw=10-20 MeV vs. the model space size and compared to the experimentally deduced values. 
    The channel radius that ensures slope continuity at matching is used. The horizontal dot-dashed lines correspond to the extrapolated (``Extrap.'') results from \hw=15 and 20 MeV calculations with the blue band representing the uncertainty.
    The calculations are for the total of $p_{1/2}$ and $p_{3/2}$ partial waves. The experimentally deduced values (``Expt.'') are from Ref. \cite{TracheACCG2003} for $\braket{^8\mathrm{Li}}{^7\mathrm{Li}+{\rm n}}$ and from Ref.
  \cite{GuoLLB2005} for $\braket{^9\mathrm{Li}}{^8\mathrm{Li}+{\rm n}}$. }
   \label{fig:8Li_9Li_ANC}
\end{figure*}

The experimentally inferred ANCs for the $\braket{^8\mathrm{Li}}{^7\mathrm{Li}+{\rm n}}$ wavefunction are available for $p_{1/2}$ and $p_{3/2}$ partial waves separately \cite{TracheACCG2003}. Thus, we compare the extrapolations of the calculated ANCs  for each of the respective partial waves (Table \ref{tab:ANC}). For this system, both \hw=15 and 20 MeV yield extrapolated results almost independent of the channel radius. For both of these HO parameters the calculations converge within the experimentally deduced range in comparatively small model spaces (Fig. \ref{fig:8Li_9Li_ANC}a). We note that only the squares of the experimentally deduced ANCs are available, and the sign of an ANC is not an observable, thus one needs to compare only the absolute values of the calculations in Table \ref{tab:ANC} to the experimentally inferred values. Nonetheless, the signs of the  ANCs, and their magnitudes, calculated from the SA-NCSM and the VMC and GFMC models are all in agreement. 

\begin{table}[tbh]
\begin{center}
\caption{\label{tab:ANC} ANCs (in fm$^{-1/2}$) from extrapolations of the SA-NCSM calculations to infinite model space and compared with the VMC, GFMC and the experimentally deduced (``Expt.'') values. Systematic uncertainties on GFMC values are 5\% or less. The experimentally deduced ANCs for the three systems are from Refs. \cite{GulamovMN1995, TracheACCG2003, GuoLLB2005}, respectively. }
\begin{tabular}{c c| c c c c } \hline \hline
Transition &  & SA-NCSM &VMC &GFMC & $|$Expt.$|$  \\ \hline
$\braket{^7\mathrm{Li}}{^6\mathrm{Li}+{\rm n}}$ &$p_{3/2}$ &1.9(1) &1.89(1) &2.29 & \\
 &$p_{1/2}$ &1.6(1) &1.65(1) &1.73 & \\
  &total & 2.5(1) &2.51(1) &2.87 &1.26 - 2.82\\
$\braket{^8\mathrm{Li}}{^7\mathrm{Li}+{\rm n}}$  &$p_{3/2}$ &$-0.72(7)$ &$-0.618(11)$ &  &$0.62(3)$\\
  &$p_{1/2}$ &0.24(4) &0.218(6) &  &$0.22(3)$\\
  &total &0.76(8) &0.655(12) &  & \\
$\braket{^9\mathrm{Li}}{^8\mathrm{Li}+{\rm n}}$  &$p_{3/2}$ &$-1.21(6)$ & $-1.140(13)$ &  & \\
  &$p_{1/2}$ &0.22(2) &0.308(7) &  & \\
   &total &1.23(6) & 1.180(15) &  &1.15(14)\\
\hline\hline
\end{tabular}
\end{center}
\end{table}

 


As mentioned above, the $\braket{^9\mathrm{Li}}{^8\mathrm{Li}+{\rm n}}$ overlap is dominated by the $p_{3/2}$ partial wave. This results in total ANC being almost indistinguishable from the $p_{3/2}$ ANC (Fig. \ref{fig:8Li_9Li_ANC}b and Table \ref{tab:ANC}). Again, the extrapolations of  ANCs for \hw=15 and 20 MeV are close to each other and practically do not depend on $r_c$. Similarly to the previous two systems, our extrapolated value agrees very well with the experimentalally deduced value from Ref. \cite{GuoLLB2005}.

 \section{Conclusions}

We have reported SA-NCSM calculations of single-neutron spectroscopic overlaps for a series of lithium isotopes using a realistic chiral potential. As expected for the HO basis, large model spaces are imperative to accommodate the tail of the overlaps, and we show that these tails converge towards the exact 
Whittaker functions
as the model space size increases. Using
these overlaps, we have calculated the associated SFs and ANCs, and showed a good agreement between them and the experimentally deduced values as well as previous GFMC and VMC calculations. The current study can be extended to heavier nuclei that are within the reach of the SA-NCSM \cite{LauneyMD_ARNPS21}. We have also discussed the effect of the normalization of the cluster wave functions, and for the illustrative example of $^6$Li+n 
the spectroscopic factors practically coincide with the measure for clustering, since the orthogonalization of the two-cluster states results in a negligible effect.


The single-nucleon overlaps can be modeled by solutions of the
Schr\"{o}dinger equation with a nucleon-nucleus effective potential.
Hence, the overlaps calculated in the \emph{ab initio} SA-NCSM approach can be used
to fit the parameters of these potentials using, \eg, Bayesian
techniques \cite{Dudeck2023prc}. This will allow one to perform uncertainty
quantification of the potential parameters. Most importantly, this will provide probability distribution functions for the parameters that will, in turn, quantify uncertainties in cross sections calculated in few-body reaction models. Such models are often employed in the analyses of experimental data, where a microscopic
input with quantified uncertainties is essential.

\section{Acknowledgements}
We thank Jutta Escher, Chlo\"e Hebborn, Gregory Potel and Konstantinos Kravvaris for useful discussions. This work was supported in part by the U.S. National Science Foundation  (PHY-1913728, PHY-2209060), the U.S. Department of Energy (DE-SC0019521, DE-SC0023532) and the Czech Science Foundation (22-14497S).  This work was performed under the auspices of the U.S. Department of Energy by Lawrence Livermore National Laboratory under Contract DE-AC52-07NA27344 and the National Nuclear Security Administration through the Center for Excellence in Nuclear Training and University Based Research (CENTAUR) under Grant No. DE-NA0003841.  This work benefited from high performance computational resources provided by LSU (www.hpc.lsu.edu),  the National Energy Research Scientific Computing Center (NERSC), a U.S. Department of Energy Office of Science User Facility operated under Contract No. DE-AC02-05CH11231, as well as the Frontera computing project at the Texas Advanced Computing Center, made possible by National Science Foundation award OAC-1818253.

\bibliographystyle{apsrev}
\bibliography{overlaps.bib, lsu_latest.bib}

\end{document}